\documentclass[aps,prx,preprint,groupedaddress,showkeys]{revtex4-2}

\usepackage{amssymb}
\usepackage{amsmath}
\usepackage{graphicx}
\usepackage{relsize}
\usepackage{verbatim}
\usepackage[utf8]{inputenc} 
\usepackage{euscript}
\usepackage{amsfonts}

\newcommand{\mds}[1] {\frac{d {#1}}{d\tau}}

\newcommand{\mdt}[1] {\frac{d {#1}}{dt}}

\begin{document}

\title{Developed Method. Interactions and their quantum picture.}
\author{Piotr Ogonowski}
\email{piotrogonowski@kozminski.edu.pl}
\affiliation{Kozminski University, Jagiellonska 57/59, Warsaw, 03-301, Poland}

\date{\today}

\begin{abstract}
By developing the previously proposed method of combining continuum mechanics with Einstein Field Equations, it has been shown that the classic relativistic description, curvilinear description, and quantum description of the physical system may be reconciled using the proposed Alena Tensor. For a system with an electromagnetic field, the Lagrangian density equal to the invariant of the electromagnetic field was obtained, vanishing four-divergence of canonical four-momentum appears to be consequence of the Poyinting theorem, and explicit form of one of gauges of the electromagnetic four-potential was introduced. The proposed method allows for further development with additional fields.
\end{abstract}

\keywords{Field theory, Lagrangian mechanics, Quantum mechanics, General Relativity, Unification of interactions}

\maketitle

\section{Introduction}

Over the past decades, great strides have been made in attempts to combine quantum description of interactions with General Relativity \cite{INT1}. There are currently many promising approaches to connecting the Quantum Mechanics and General Relativity, including perhaps the most promising ones: Loop Quantum Gravity \cite{LQG1}, \cite{LQG2}, \cite{LQG3}, String Theory \cite{ST1}, \cite{ST2}, \cite{ST3} and Noncommutative Spacetime Theory \cite{NS1}, \cite{NS2}. \\
~\\
There are also attempts to modify General Relativity or find an equally good alternative theory \cite{ALT1}, \cite{ALT2}, \cite{ALT3} that would provide a more general description or would allow for the inclusion of other interactions. A lot of work has also been done to clear up some challenges related to General Relativity and $\Lambda CMD$ model \cite{LCMD}. An explanation for the problem of dark energy \cite{DE1} and dark matter \cite{DM1} is still being sought, and efforts are still being made to explain the origin of the cosmological constant  \cite{CC1}, \cite{CC2}, \cite{CC3}. \\
~\\
The author also tries to bring his own contribution to the explanation of the above physics challenges, based on a recently discovered method, described in \cite{PMofC}. As this article will show, this method seems very promising and may help clarify at least some of the issues mentioned above. The author's method, similar to the approach presented in \cite{DFF}, \cite{LLM}, \cite{GEM1} also points to the essential connections between electromagnetism and Gereral Relativity, however, the postulated relationship is of a different nature and can be percieved as some generalization of the direction of research proposed in \cite{BCIJ}, \cite{VGRA}, \cite{PGRA} and \cite{AGRA}.\\
~\\
According to conclusions from \cite{PMofC}, the description of motion in curved spacetime and its description in flat Minkowski spacetime with fields are equivalent, and the transformation between curved spacetime and Minkowski spacetime is known, because the geometry of curved spacetime depends on the field tensor. This allows for a significant simplification of research, because the results obtained in flat Minkowski spacetime can be easily transformed into curved spacetime. The last missing link seems to be the quantum description.\\
~\\
In this article, the author will focus on developing the proposed method for a system with electromagnetic field in such a way, as to obtain the convergence with the description of QED and Quantum Mechanics. In the first chapter, the Lagrangian density for the system will be derived, allowing to obtain the tensor described in \cite{PMofC}. These conclusions will be used later in the article to propose possible directions of research on combining the GR description with QFT and QM.\\
~\\ 
The author uses the Einstein summation convention, metric signature $(+,-,-,-)$ and commonly used notations. In order to facilitate the analysis of the article, the key conclusions from \cite{PMofC} are quoted in the subsection below. 

\subsection{Short summary of the method} \label{ShortSummary}

According to \cite{PMofC}, stress-energy tensor $T^{\alpha \beta}$ for a system with electromagnetic field in a given spacetime, described by a metric tensor $g^{\alpha \beta}$ is equal to
\begin{equation} \label{CH2}
T^{\alpha \beta} =  \varrho \, U^{\alpha} U^{\beta} - \left ( c^2 \varrho +  \Lambda_{\rho}  \right ) \left ( g^{\alpha \beta} - \xi \, h^{\alpha \beta} \right )
\end{equation}
where $\varrho_o$ is for rest mass density, $\gamma$ is Lorentz gamma factor, and
\begin{equation} \label{VRD}
\varrho \equiv \varrho_o \gamma
\end{equation}
\begin{equation} \label{CH4}
\frac{1}{\xi} \equiv \frac{1}{4}\, g_{\mu \nu} \, h^{\mu \nu}
\end{equation}
\begin{equation} \label{CH5}
\Lambda_{\rho} \equiv \frac{1}{4\mu_o} \mathbb{F}^{\alpha \mu} \, g_{\mu \gamma} \, \mathbb{F}^{ \beta \gamma } g_{\alpha \beta}
\end{equation}
\begin{equation} \label{CH6}
h^{\alpha \beta} \equiv 2 \; \frac{\mathbb{F}^{\alpha \delta} \, g_{\delta \gamma} \, \mathbb{F}^{ \beta \gamma } }{ \sqrt{\mathbb{F}^{\alpha \delta} \, g_{\delta \gamma} \, \mathbb{F}^{ \beta \gamma } \, g_{\mu \beta} \, \mathbb{F}_{\alpha \eta} \, g^{\eta \xi} \, \mathbb{F}^{\mu}_{\; \; \xi }}}
\end{equation}
In above, $\mathbb{F}^{\alpha \beta}$ represents electromagnetic field tensor, and the stress–energy tensor for electromagnetic filed, denoted as $\Upsilon^{\alpha \beta}$ may be presented as follows
\begin{equation} \label{CH7}
\Upsilon^{\alpha \beta} \equiv  \Lambda_{\rho} \left (g^{\alpha \beta} - \xi \, h^{\alpha \beta}  \right ) = \Lambda_{\rho} g^{\alpha \beta} - \frac{1}{\mu_o}\mathbb{F}^{\alpha \delta} \, g_{\delta \gamma} \, \mathbb{F}^{ \beta \gamma }
\end{equation}
Thanks to the proposed amendment to the continuum mechanics, in flat Minkowski spacetime occurs
\begin{equation} \label{CH1}
\partial_{\alpha} U^{\alpha} = -\mdt{\gamma}  \quad \to \quad \partial_{\alpha} \, \varrho U^{\alpha} = 0
\end{equation}
thus denoting four-momentum density as $\varrho U^{\mu} = \varrho_o \gamma \, U^{\mu}$, total four-force density $f^{\mu}$ acting in the system is
\begin{equation} \label{TotFoI}
f^{\mu} \equiv \varrho A^{\mu} = \partial_{\alpha} \varrho U^{\mu} U^{\alpha}
\end{equation}
Denoting rest charge density in the system as $ \rho_o$ and
\begin{equation} \label{RHZERO}
\rho \equiv \rho_o \gamma
\end{equation}
electromagnetic four-current $J^{\alpha}$ is equal to
\begin{equation} \label{EN11}
J^{\alpha} \equiv \rho \, U^{\alpha} = \rho_o \gamma \, U^{\alpha}
\end{equation}
The pressure p in the system is equal to
\begin{equation} \label{Pressure}
p \equiv c^2 \varrho +  \Lambda_{\rho}
\end{equation}
In the flat Minkowski spacetime, total four-force density $f^{\alpha}$ acting in the system calculated from vanishing $\partial_{\beta} \; T^{\alpha \beta}$ is the sum of electromagnetic ($f^{\alpha}_{EM}$), gravitational ($f^{\alpha}_{gr}$) and the sum of remaining ($f^{\alpha}_{oth}$) four-force densities
\begin{equation} \label{forceDen}
f^{\alpha} = 
\begin{cases} 
f^{\alpha}_{EM} \equiv \partial_{\beta} \; \Upsilon^{\alpha \beta} \quad (electromagnetic) \\
+ \\
f^{\alpha}_{gr} \equiv \left ( g^{\alpha \beta} - \xi \, h^{\alpha \beta}  \right ) \partial_{\beta} \, p \quad (gravitational) \\
+ \\
f^{\alpha}_{oth} \equiv \frac{\varrho c^2}{\Lambda_{\rho}} f^{\alpha}_{EM}  \quad (sum \, of\, remaining \, forces)
\end{cases}
\end{equation}
As was shown in \cite{PMofC}, in curved spacetime ($g_{\alpha \beta} =h_{\alpha \beta}$) presented method reproduces Einstein Field Equations with an accuracy of $\frac{4\pi G}{c^4}$ constant and with cosmological constant $\Lambda$ dependent on invariant of electromagnetic field tensor $\mathbb{F}^{\alpha \gamma}$
\begin{equation} \label{CosmoCon}
\Lambda =  - \frac{\pi G}{c^4 \mu_o} \, \mathbb{F}^{\alpha \mu} \, h_{\mu \gamma} \, \mathbb{F}^{ \beta \gamma } h_{\alpha \beta} = - \frac{4\pi G}{c^4} \Lambda_{\rho}
\end{equation}
where $h_{\alpha \beta}$ appears to be metric tensor of the spacetime in which all motion occurs along geodesics and where $\Lambda_{\rho}$ describes vacuum energy density. \\
~\\
It is worth noting that although in flat Minkowski spacetime $\Lambda_{\rho}$ has a negative value due to the adopted metric signature, this does not determine its value in curved spacetime. Therefore, solutions with a negative cosmological constant are also possible, which is an issue discussed in the literature \cite{NL1}, \cite{NL2}, \cite{NL3}.\\
~\\
It was also shown, that in this solution, Einstein tensor describes the spacetime curvature related to vanishing in curved spacetime four-force densities $f^{\alpha}_{gr} + f^{\alpha}_{oth}$ and is therefore related to the curvature responsible for gravity only when other forces are neglected.\\
~\\
The proposed method allows to add additional fields while maintaining its properties. One may define another stress-energy tensor describing the field (e.g. describing the sum of several fields) instead of $\Upsilon^{\alpha \beta}$ and insert it into the stress-energy tensor $T^{\alpha \beta}$ in a manner analogous to that presented above. As a result of the vanishing four-divergence of $T^{\alpha \beta}$ one will obtain in flat spacetime four-force densities related to the new field, and in curved spacetime, the equations will transform into EFE with the cosmological constant depending on the invariant of the considered, new field strength tensor.

\section{Lagrangian density for the system} \label{SepOfFields}

Since for the considered method the transition to curved spacetime is known (based on electromagnetic field tensor), the rest of the article will focus on the calculations in the Minkowski spacetime with presence of electromagnetic field, where $\eta^{\alpha \beta}$ represents Minkowski metric tensor. \\
~\\
Using a simplified notation
\begin{equation} \label{simplfNot}
\mds{\, \ln{(p)}} = U_{\mu} \partial^{\mu} \ln{(p)} = U_{\mu} \frac{\partial^{\mu} p}{p}
\end{equation}
it can be seen that the four-force densities resulting from the obtained stress-energy tensor (\ref{forceDen}) in flat Minkowski spacetime can be written as follows
\begin{equation} \label{forceDenNew}
\begin{cases} 
f^{\alpha}_{gr} = \left ( \eta^{\alpha \beta} - \xi \, h^{\alpha \beta}  \right ) \partial_{\beta} p = \mds{\,\ln{(p)}} \varrho U^{\alpha} - T^{\alpha \beta}  \partial_{\beta} \ln{(p)}  \\
f^{\alpha}_{EM} = \frac{\Lambda_{\rho}}{p} \left ( f^{\alpha} - f^{\alpha}_{gr} \right )\\
f^{\alpha}_{oth} = \frac{\varrho c^2}{p} \left ( f^{\alpha} - f^{\alpha}_{gr} \right )
\end{cases}
\end{equation}
where $f^{\mu}_{EM}$ can also be represented in terms of electromagnetic four-potential and four-current. This means that to fully describe the system and derive the Lagrangian density, it is enough to find an explicit equation for the gravitational force or some gauge of electromagnetic four potential.\\
~\\
Referring to definitions from section \ref{ShortSummary} one may notice, that by proposing following electromagnetic four-potential $\mathbb{A}^{\mu}$
\begin{equation} \label{EM4pot}
\mathbb{A}^{\mu} \equiv - \frac{\Lambda_{\rho}}{p} \frac{\varrho_o}{\rho_o} U^{\mu}
\end{equation}
one obtains electromagnetic four-force density $f^{\alpha}_{EM}$ in form of
\begin{equation} \label{EM4EMf}
f^{\alpha}_{EM} = J_{\beta} \left ( \partial^{\alpha}\mathbb{A}^{\beta} - \partial^{\beta}\mathbb{A}^{\alpha} \right ) = \frac{\Lambda_{\rho}}{p} \left ( f^{\alpha} - \mds{\,\ln{(p)}} \varrho U^{\alpha} + \varrho c^2 \partial^{\alpha} \ln{(p)}  \right )
\end{equation}
where $J_{\beta}$ is electromagnetic four-current and where Minkowski metric property was utilized
\begin{equation} \label{MPR}
U_{\beta}U^{\beta} = c^2 \quad \rightarrow \quad U_{\beta}\partial^{\alpha}U^{\beta}=\frac{1}{2}\;  \partial^{\alpha} \left ( U_{\beta}U^{\beta} \right ) = 0 
\end{equation}
Four-force densities acting in the system may be now described by the following equality
\begin{equation} \label{Equality}
J_{\beta} \left ( \partial^{\alpha}\mathbb{A}^{\beta} - \partial^{\beta}\mathbb{A}^{\alpha} \right ) + \varrho U_{\beta} \left ( \partial^{\beta} \frac{\varrho c^2}{p} U^{\alpha} - \partial^{\alpha}\frac{\varrho c^2}{p} U^{\beta} \right ) = \varrho U_{\beta} \left ( \partial^{\beta} U^{\alpha} - \partial^{\alpha} U^{\beta} \right ) = f^{\alpha}
\end{equation}
Comparing (\ref{forceDenNew}) and (\ref{EM4EMf}) it is seen, that introduced electromagnetic four-potential yields 
\begin{equation} \label{For0tens}
0 = \left ( T^{\alpha \beta} - \varrho c^2 \; \eta^{\alpha \beta} \right ) \, \partial_{\beta} \ln{(p)}
\end{equation}
which is equivalent to imposing following condition on normalized stress-energy tensor
\begin{equation} \label{normCond}
0 = \partial_{\beta} \left ( \frac{T^{\alpha \beta}}{\eta_{\mu \gamma}T^{\mu \gamma}}  \right ) + \partial^{\alpha} \ln{\left ( \eta_{\mu \gamma} T^{\mu \gamma} \right )}
\end{equation}
and what yields gravitational four-force density in Minkowski spacetime in form of
\begin{equation} \label{SIMGR}
f^{\mu}_{gr} = \varrho \left ( \mds{\, \ln{(p)}} U^{\mu} - c^2 \partial^{\mu} \ln{(p)} \right )
\end{equation}
Now, one may show, that proposed electromagnetic four-potential leads to correct solutions.\\
~\\
At first, recalling the classical Lagrangian density \cite{LagDen} for electromagnetism one may show why, in the light of the conclusions from \cite{PMofC} and above, it does not seem to be correct and thus makes it difficult to create a symmetric stress-energy tensor \cite{MEDP}. The classical value of the Lagrangian density for electromagnetic field, written with the notation used in the article, is
\begin{equation} \label{CLASSLAG}
- \mathcal{L}_{EMclassic} = \Lambda_{\rho} + \mathbb{A}^{\mu} J_{\mu}
\end{equation}
In addition to the obvious doubt that by taking the different gauge of the four-potential $\mathbb{A}^{\mu}$ one changes the value of the Lagrangian density, one may notice, that with considered electromagnetic four-potential, such Lagrangian density is equal to
\begin{equation} \label{CLASSLAG2}
- \mathcal{L}_{EMclassic} = \Lambda_{\rho} - \frac{\Lambda_{\rho}}{p} \frac{\varrho_o}{\rho_o} U^{\mu} \, U_{\mu} \rho_o \gamma = \Lambda_{\rho} - \frac{\Lambda_{\rho} \varrho c^2}{p} = \frac{\Lambda_{\rho}^2}{p}
\end{equation}
As it is seen, above Lagrangian density is not invariant under gradient over four-position and $\mathbb{A}^{\mu}$ and $J_{\mu}$ are dependent, what is not taken into account in classical calculation
\begin{equation} \label{DEPENDFIELDS}
\frac{\mathbb{A}^{\alpha}}{\mathbb{A}^{\mu} \mathbb{A}_{\mu}}  = \frac{J^{\alpha}}{\mathbb{A}^{\mu} J_{\mu}} 
\end{equation}
Above yields 
\begin{equation} \label{GRADPOT}
\frac{\partial \ln{\left ( \frac{1}{\sqrt{\mathbb{A}^{\mu} \mathbb{A}_{\mu}}} \right)}}{\partial \mathbb{A}_{\alpha}} = - \frac{J^{\alpha}}{\mathbb{A}^{\mu} J_{\mu}} = \frac{p}{\varrho c^2} \frac{J^{\alpha}}{\Lambda_{\rho}}
\end{equation}
One may decompose
\begin{equation} \label{DECPOT}
\ln{\left ( \frac{1}{\sqrt{\mathbb{A}^{\mu} \mathbb{A}_{\mu}}} \right)} = \ln{\left ( \frac{p \, \rho_o }{ \varrho_o c} \right)} - \ln{\left (\Lambda_{\rho} \right ) }  
\end{equation}
and simplify (\ref{GRADPOT}) to
\begin{equation} \label{GRAFIN}
\frac{\partial \ln{\left ( \frac{p \, \rho_o }{ \varrho_o c} \right)} }{\partial \mathbb{A}_{\alpha}}  - \frac{\partial \ln{\left (\Lambda_{\rho} \right ) } }{\partial \mathbb{A}_{\alpha}} = \frac{J^{\alpha}}{\varrho c^2} + \frac{J^{\alpha}}{\Lambda_{\rho}}
\end{equation}
Above yields
\begin{equation} \label{MUINV}
\frac{\partial \ln{\left ( \Lambda_{\rho} \right ) }}{\partial \mathbb{A}_{\alpha}} = - \frac{J^{\alpha}}{\Lambda_{\rho}}
\end{equation}
which leads to the conclusion that $\Lambda_{\rho}$ acts as the Lagrangian density for the system
\begin{equation} \label{CLASSLAGNEW}
\frac{\partial \Lambda_{\rho}} {\partial \mathbb{A}_{\alpha}} =  \partial_{\nu}  \left (  \frac{\partial \Lambda_{\rho}}{\partial (\partial_{\nu} \mathbb{A}_{\alpha} )}   \right ) = - J^{\alpha}
\end{equation}
which would support conclusions from \cite{ALL1} and what yields stress-energy tensor for the system in form of
\begin{equation} \label{TensSep1}
T^{\alpha \beta} = \frac{1}{\mu_o}\mathbb{F}^{\alpha \gamma} \partial^{\beta} \mathbb{A}_{\gamma} -  \Lambda_{\rho} \eta^{\alpha \beta}
\end{equation}
In fact, the proof of correctness of the electromagnetic field tensor (noted as $\Upsilon^{\alpha \beta}$) allows to see this solution
\begin{equation} \label{Ups1}
f_{EM}^{\beta} = \partial_{\alpha} \Upsilon^{\alpha \beta} = J^{\gamma}\mathbb{F}^{\beta}_{\, \, \, \gamma} - \frac{1}{\mu_o}\mathbb{F}^{\alpha \gamma} \partial_{\alpha} \mathbb{F}^{\beta}_{\, \, \, \gamma}
\end{equation}
what yields following property of electromagnetic field tensor
\begin{equation} \label{Ups2}
\mathbb{F}^{\alpha \gamma} \partial_{\alpha} \partial_{\gamma} \mathbb{A}^{\beta} 
 = \mathbb{F}^{\alpha \gamma}  \partial^{\beta} \partial_{\alpha} \mathbb{A}_{\gamma} \end{equation}
Using the above substitution, one may note
\begin{equation} \label{TensorProof}
\partial_{\alpha} \Upsilon^{\alpha \beta} = \partial_{\alpha} \frac{1}{\mu_o} \mathbb{F}^{\alpha \gamma} \partial_{\gamma} \mathbb{A}^{\beta} - \partial_{\alpha} \frac{1}{\mu_o} \mathbb{F}^{\alpha \gamma} \partial^{\beta} \mathbb{A}_{\gamma} = \frac{1}{\mu_o} \mathbb{F}^{\alpha \gamma} \partial^{\beta} \partial_{\alpha} \mathbb{A}_{\gamma} - J^{\gamma}\partial_{\gamma} \mathbb{A}^{\beta} - \partial_{\alpha} \frac{1}{\mu_o} \mathbb{F}^{\alpha \gamma} \partial^{\beta} \mathbb{A}_{\gamma}
\end{equation}
Therefore the invariance of $\Lambda_{\rho}$ with respect to $\mathbb{A}_{\alpha}$ and  $\partial_{\nu} \mathbb{A}_{\alpha}$ is both a condition on the correctness of the electromagnetic stress-energy tensor and on $\Lambda_{\rho}$ in the role of Lagrangian density
\begin{equation} \label{TENSCORR}
0 = \partial^{\beta} \Lambda_{\rho} = \frac{\partial \Lambda_{\rho}}{\partial (\partial_{\alpha} \mathbb{A}_{\gamma})} \partial^{\beta} (\partial_{\alpha} \mathbb{A}_{\gamma}) + \frac{\partial \Lambda_{\rho}} {\partial \mathbb{A}_{\gamma}} \partial^{\beta} \mathbb{A}_{\gamma} = \frac{1}{\mu_o} \mathbb{F}^{\alpha \gamma} \partial^{\beta} \partial_{\alpha} \mathbb{A}_{\gamma}  - J^{\gamma} \partial^{\beta} \mathbb{A}_{\gamma} = \partial_{\alpha} \frac{1}{\mu_o} \mathbb{F}^{\alpha \gamma} \partial^{\beta} \mathbb{A}_{\gamma}
\end{equation}
what yields for (\ref{TensorProof}) that
\begin{equation} \label{UpsFin}
\partial_{\alpha} \Upsilon^{\alpha \beta} = J^{\gamma} \partial^{\beta} \mathbb{A}_{\gamma} - J^{\gamma}\partial_{\gamma} \mathbb{A}^{\beta} = f_{EM}^{\beta}
\end{equation}
Equations (\ref{CH2}), (\ref{CH7}) and (\ref{TensSep1}) yield
\begin{equation} \label{TensSep2}
\frac{1}{\mu_o} \mathbb{F}^{\alpha \gamma} \partial_{\gamma} \mathbb{A}^{\beta} = \varrho \, U^{\alpha} U^{\beta} - \frac{c^2 \varrho }{\Lambda_{\rho}} \, \Upsilon^{\alpha \beta}
\end{equation}
what yields second representation of the stress-energy tensor
\begin{equation} \label{TensSep3}
T^{\alpha \beta} = \frac{p}{\varrho c^2} \cdot \frac{1}{\mu_o} \mathbb{F}^{\alpha \gamma} \partial_{\gamma} \mathbb{A}^{\beta} - \frac {\Lambda_{\rho}}{c^2} U^{\alpha} U^{\beta} = \frac{p}{\varrho c^2} \, \partial_{\gamma} \frac{1}{\mu_o} \mathbb{F}^{\alpha \gamma} \mathbb{A}^{\beta}
\end{equation}
After four-divergence, it gives additional expression for relation between forces and gives useful clues about the behavior of the system when transitioning to the description in curved spacetime.

\section{Hamiltonian density and energy transmission} \label{Hamilton}

Noting Hamiltonian density as $\mathcal{H}$, from (\ref{TensSep1}) one gets
\begin{equation} \label{HQ2}
\mathcal{H} \equiv  T^{0 0} = \frac{1}{\mu_o}\mathbb{F}^{0 \gamma} \partial^{0} \mathbb{A}_{\gamma} - \Lambda_{\rho}
\end{equation}
Above Hamiltonian density agrees with the classical Hamiltonian density for electromagnetic field \cite{LDEM} except that this Hamiltonian density was currently mainly considered for sourceless regions. According to conclusions from previous chapter, this Hamiltonian density describes the whole system with electromagnetic field, including gravity and other four-force densities resulting from considered stress-energy tensor. Above, therefore, may significantly simplify Quantum Field Theory equations  \cite{QFT1}, \cite{QFT2}, \cite{QFT3} which will be shown in this chapter for the purposes of QED.\\
~\\
At first one may notice, that in transformed (\ref{TensSep1})
\begin{equation} \label{HQ4}
- T^{\alpha 0} =  - \frac{1}{\mu_o} \mathbb{F}^{\alpha \gamma} \partial_{\gamma} \mathbb{A}^{0} + \Upsilon^{\alpha 0}
\end{equation}
first row of electromagnetic stress-energy tensor $\Upsilon^{\alpha 0}$ is a four-vector, representing energy density of electromagnetic field and Poynting vector \cite{PoV} - the Poynting four-vector. Therefore vanishing four-divergence of the $T^{\alpha 0}$ must represent Poynting theorem. Indeed, properties (\ref{Ups2}) and (\ref{TENSCORR}) provide such equality
\begin{equation} \label{HQ5}
0 = - \partial_{\alpha} T^{\alpha 0} = J_{\gamma} \mathbb{F}^{0 \gamma} + \partial_{\alpha} \Upsilon^{\alpha 0} 
\end{equation}
~\\
Next, one may introduce auxiliary variable $\varepsilon$ with the energy density dimension, defined as follows
\begin{equation} \label{EpsDef}
c\varepsilon \equiv - \frac{1}{\mu_o} \mathbb{F}^{0 \mu} \mds {\mathbb{A}_{\mu}}
\end{equation}
and comparing the result
\begin{equation} \label{Lagr}
- U_{\beta}T^{0 \beta} = c\varepsilon + c\gamma \Lambda_{\rho}
\end{equation}
between the two tensor definitions (\ref{TensSep1}), (\ref{TensSep3}) one may notice, that it must hold
\begin{equation} \label{EpsProp}
- \frac{p}{\varrho c^2} \cdot \frac{1}{\mu_o} \mathbb{F}^{0 \mu} \partial_{\mu} \mathbb{A}^{\beta} = - \frac{p}{\varrho c^2 \mu_o} \cdot \left ( U^{\beta} \, \mathbb{F}^{0 \mu} \partial_{\mu} \frac{\mathbb{A}^0}{c\gamma} + \frac{\mathbb{A}^0}{c\gamma} \mathbb{F}^{0 \mu} \partial_{\mu} U^{\beta} \right )  = \frac{\varepsilon}{c} \, U^{\beta} - \frac{p}{\varrho c^2} \frac{\mathbb{A}^0}{\gamma c\mu_o} \mathbb{F}^{0 \mu} \partial_{\mu} U^{\beta}
\end{equation}
because the second component of above vanishes contracted with $U_{\beta}$, due to the property of the Minkowski metric (\ref{MPR}). Therefore (\ref{TensSep1}) and (\ref{TensSep3}) also yield
\begin{equation} \label{TensSepTrans}
- T^{0 \beta} =  \varepsilon \frac{\varrho c}{p} \, U^{\beta} - \frac{c \epsilon_o \mathbb{A}^0}{\gamma} \mathbb{F}^{0 \mu} \partial_{\mu} U^{\beta} + \Upsilon^{0 \beta}
\end{equation}
where $\epsilon_o$ is electric vacuum permittivity, and
\begin{equation} \label{YUeq}
\Upsilon^{0 \beta} U_{\beta} = c\varepsilon \frac{\Lambda_{\rho}}{p} + c\gamma \Lambda_{\rho}
\end{equation}
Since $\partial^{\mu} p = \partial^{\mu} \varrho c^2$ thus from (\ref{EpsProp}) one gets
\begin{equation} \label{Epsilon}
\varepsilon \gamma = \frac{ c \epsilon_o \mathbb{A}^0}{\gamma} \; \mathbb{F}^{0 \mu} \partial_{\mu} \gamma c 
\end{equation}
and thanks to (\ref{EpsProp}) substituted to (\ref{TensSep3}) one also obtains
\begin{equation} \label{TenyL}
- T^{0 \beta} =  \frac{\varepsilon + \Lambda_{\rho} \gamma}{c} \, U^{\beta} - \frac{p}{\varrho c^2} \frac{c \epsilon_o \mathbb{A}^0}{\gamma } \mathbb{F}^{0 \mu} \partial_{\mu} U^{\beta} 
\end{equation}
Since from (\ref{CH2}) and (\ref{CH7}) for $T^{0 0}$ one gets
\begin{equation} \label{Ten0Com}
\mathcal{H} = \varrho c^2 \gamma^2 - \frac{p}{\Lambda_{\rho}}  \Upsilon^{0 0}
\end{equation}
therefore comparing zero-component of (\ref{TensSepTrans}) 
\begin{equation} \label{0Component1}
\mathcal{H} = - \varepsilon \gamma \frac{\varrho c^2}{p} + \varepsilon \gamma - \Upsilon^{0 0} = \varepsilon \gamma - \frac{\mathcal{H}}{\Lambda_{\rho}} \varrho c^2 - \frac{p}{\Lambda_{\rho}}  \Upsilon^{0 0}
\end{equation}
to (\ref{Ten0Com}) and comparing to (\ref{TenyL}) 
\begin{equation} \label{0Component2}
\mathcal{H} = - \varepsilon \gamma - \Lambda_{\rho} \gamma^2 + \frac{p}{\varrho c^2} \varepsilon \gamma = \Lambda_{\rho} \left ( \frac{\varepsilon \gamma}{\varrho c^2} - \gamma^2  \right )
\end{equation}
one may notice, that
\begin{equation} \label{HQ4e}
\varepsilon \gamma = \varrho c^2 \gamma ^2 + \varrho c^2
\end{equation}
is a valid solution of the system, what yields
\begin{equation} \label{Hvalue}
\mathcal{H} = \Lambda_{\rho}
\end{equation}
\begin{equation} \label{LDR}
\frac{1}{c\gamma} U_{\beta} T^{0 \beta} = - \frac{\varrho_o c^2}{\gamma} - p
\end{equation} 
In fact, there is a whole class of solutions (\ref{HQ4e}) in the form $\varepsilon \gamma = \varrho c^2 \gamma ^2 + \mathcal{K} \cdot \varrho c^2$, however, $\mathcal{K}<>1$ would not be consistent with the following conclusions. It is also worth noting that the obtained solution $\mathcal{H} = \mathcal{L} = \Lambda_{\rho}$ means, that there is no potential in the system in the classical sense, thus the dynamics of the system depends on itself. This is exactly what would be expected from a description that reproduces General Relativity in flat spacetime. \\
~\\
From the analysis of the equation (\ref{TensSepTrans}) it may be then concluded, that after integration of the $-\frac{1}{c}T^{0 \beta}$ with respect to the volume, the total energy transported in the isolated system should be the sum of the four-momentum and four-vectors describing energy transmission related to fields. This would be consistent with the conclusion from \cite{SETH} that "equations of motion for matter do not need to be introduced separately, but follow from the field equations". It would mean, that the canonical four-momentum density is just a part of the stress-energy tensor.\\
~\\
Therefore, by analogy with the Poyting four-vector $\frac{1}{c}\Upsilon^{0 \beta}$, one may introduce a four-vector $Z^{\beta}$ understood as its equivalent for the remaining interactions and rewrite (\ref{TensSepTrans}) as
\begin{equation} \label{HQ4b}
- \frac{1}{c} T^{0 \beta} =  \varrho_o U^{\beta} + Z^{\beta} + \frac{1}{c}\Upsilon^{0 \beta}
\end{equation}
where
\begin{equation} \label{HQ4f}
Z^{\beta} \equiv \rho_o \mathbb{A}^{\beta} + \frac{\varrho c^2 \gamma^2}{p}\varrho_o U^{\beta} - \frac{\epsilon_o \mathbb{A}^0}{\gamma} \mathbb{F}^{0 \mu} \partial_{\mu} U^{\beta}
\end{equation}
The above result ensures that the canonical four-momentum density for the system with electromagnetic field depends on the four-potential and charge density as expected. This supports the earlier statement about the need to set $\mathcal{K}=1$ and makes its physical interpretation visible. It is also worth to notice, that $ -\frac{\epsilon_o \mathbb{A}^0}{\gamma} \mathbb{F}^{0 \mu} \partial_{\mu} U^{\beta}$, due to its properties, may be associated with some description of the spin.\\
~\\
One may also note, that the above solution yields $p < 0$ since energy density of electromagnetic field is
\begin{equation} \label{HQ4g}
\Upsilon^{0 0} = \frac{\Lambda_{\rho}}{p} \left (  \varrho c^2 \gamma ^2 - \Lambda_{\rho} \right )
\end{equation}
where $\Lambda_{\rho} < 0$ in flat spacetime, due to the adopted metric signature. Thus $Z^{\beta}$ may also be simplified to
\begin{equation} \label{HQ4h}
Z^{\beta} = \frac{\Upsilon^{0 0}}{\Lambda_{\rho}} \varrho_o U^{\beta} - \frac{\epsilon_o \mathbb{A}^0}{\gamma} \mathbb{F}^{0 \mu} \partial_{\mu} U^{\beta}
\end{equation}
~\\
Finally, one may define another gauge $\bar{\mathbb{A}}_{\gamma}$ of electromagnetic four-potential $\mathbb{A}_{\gamma}$ in following way
\begin{equation} \label{HQ6}
\bar{\mathbb{A}}^{\gamma} \equiv  \mathbb{A}^{\gamma} - \partial^{\gamma} \mathbb{A}^{\beta}X_{\beta} = - X_{\beta} \partial^{\gamma} \mathbb{A}^{\beta}
\end{equation}
and note, that
\begin{equation} \label{HQ7}
- X_{\beta} T^{0 \beta} =  \frac{1}{\mu_o} \mathbb{F}^{0 \gamma} \bar{\mathbb{A}}_{\gamma} + X_{\beta} \Upsilon^{0 \beta}
\end{equation}
Four-divergence of $T^{0 \beta}$ vanishes, therefore (\ref{Hvalue}) indicates that
\begin{equation} \label{HQ8}
X_{\beta} \partial^{\alpha} T^{0 \beta} = 0
\end{equation}
what yields
\begin{equation} \label{HQ9}
\partial^{\alpha} X_{\beta} T^{0 \beta} =  T^{0 \alpha}
\end{equation}
Above brings two more important insights: 
\begin{itemize}
  \item $\frac{1}{c}X_{\beta} T^{0 \beta}$ may play a role of the density of Hamilton's principal function,
  \item Hamilton's principal function may be expressed based on the electromagnetic field only, so in the absence of the electromagnetic field it disappears.
\end{itemize}
All above also drives to conclusion, that (\ref{LDR}) may also act as Lagrangian density, used in the classic relativistic description based on four-vectors. \\
~\\
One may thus summarize all of the above findings and propose a method for the description of the system with the use of classical field theory for point-like particles. 

\section{Point-like particles and their quantum picture} \label{Quantas}

To begin with, it should be noted that the reasoning presented in chapter \ref{Hamilton} changes the interpretation of what the relativistic principle of least action means. As one may conclude from above, there is no inertial system in which no fields act, and in the absence of fields, the Lagrangian, the Hamiltonian and Hamilton's principal function vanish. Since the metric tensor (\ref{CH6}) for description in curved spacetime depends on the electromagnetic field tensor only, it seems clear, that in the considered system, the absence of the electromagnetic field means actually the disappearance of spacetime and the absence of any action.\\
~\\
One may then introduce generalized, canonical four-momentum $H^{\mu}$ as four-gradient on Hamilton's principal function S
\begin{equation} \label{Q2}
H^{\mu} \equiv - \frac{1}{c} \int T^{0 \mu} \, d^3 x \equiv -\partial^{\mu} S
\end{equation}
where 
\begin{equation} \label{Q2a}
- S \equiv H^{\mu}X_{\mu}
\end{equation}
One may also conclude from previous chapter, that canonical four-momentum should be in form of
\begin{equation} \label{Q5}
H^{\mu} = P^{\mu} + V^{\mu} 
\end{equation}
where 
\begin{equation} \label{CD1}
V^{\mu} \equiv \int Z^{\mu} + \frac{1}{c} \Upsilon^{0 \mu} \, d^3 x
\end{equation}
and where four-momentum $P^{\mu}$ may be now considered as just "other gauge" of  $-V^{\mu}$ 
\begin{equation} \label{Q5b}
- \partial^{\alpha} P^{\mu} = \partial^{\alpha} \partial^{\mu} S + \partial^{\alpha} V^{\mu}
\end{equation}
Since in the limit of the inertial system one gets $P^{\mu} X_{\mu} = mc^2\tau$, therefore, to ensure vanishing Hamilton's principal function in the inertial system, one may expect that
\begin{equation} \label{Q1}
V^{\mu} X_{\mu} \equiv - mc^2 \tau 
\end{equation}
what would also yield vanishing in the inertial system Lagrangian $L$ for point-like particle in form of
\begin{equation} \label{Q4}
- \gamma L = U_{\mu} H^{\mu}= F^{\mu} X_{\mu} 
\end{equation}
where $F^{\mu}$ is four-force. Eq. (\ref{TenyL}) yields
\begin{equation} \label{Q5}
H^{\mu} = - \frac{\gamma L}{c^2}U^{\mu} + \mathbb{S}^{\mu}
\end{equation}
where
\begin{equation} \label{Q5}
\mathbb{S}^{\beta} \equiv \int \frac{\epsilon_o \Lambda_{\rho}}{\gamma c \rho_o} \; \mathbb{F}^{0 \mu} \partial_{\mu} U^{\beta}  \, d^3 x
\end{equation}
and where $\mathbb{S}^{\beta}U_{\beta}$ vanishes, what comes from 
\begin{equation} \label{Q5a}
\mathbb{S}^{\beta} = \mds{S}  \frac{1}{c^2} U^{\beta} - \partial^{\beta} S 
\end{equation}
In above picture, the Hamilton's principal function, generalized canonical four-momentum and Lagrangian vanish for inertial system as expected. \\
~\\
Since
\begin{equation} \label{Q6}
\mathbb{S}^{\mu}\mathbb{S}_{\mu} = H^{\mu}H_{\mu} - \left ( \frac{\gamma L}{c} \right )^2
\end{equation}
therefore, to ensure compatibility with the equations of quantum mechanics it suffices to consider properties of $\mathbb{S}^{\mu}\mathbb{S}_{\mu}$. For instance, if
\begin{equation} \label{Q7}
\mathbb{S}^{\mu}\mathbb{S}_{\mu} = m^2c^2 - \left ( \frac{\gamma L}{c} \right )^2
\end{equation}
then, by introducing quantum wave function $\Psi$ in form of 
\begin{equation} \label{Q8}
\Psi \equiv e^{\pm iK^{\mu} X_{\mu}}
\end{equation} 
where $K^{\mu}$ is wave four-vector related to canonical four-momentum 
\begin{equation} \label{Q9}
\hbar  K^{\mu} \equiv H^{\mu}
\end{equation}
from (\ref{Q6}) one obtains Klein-Gordon equation
\begin{equation} \label{Q10}
\left ( \Box + \frac{m^2c^2}{\hbar^2} \right ) \Psi = 0
\end{equation} 
This shows that in addition to aligment with QFT (\ref{HQ2}), the first quantization also seems possible which allows for further analysis of the system from the perspective of the Quantum Mechanics, eliminating the problem of negative energy appearing in solutions \cite{KGE}. \\
~\\
The above representation allows the analysis of the system in the quantum approach, classical approach based on (\ref{HQ4}) and the introduction of a field-dependent metric in (\ref{CH6}) for curved spacetime, which connects previously divergent descriptions of physical systems.

\section{Conclusions and Discussion} \label{3Chap}

As shown above, the proposed method summarized in section \ref{ShortSummary} seems to be very promising area of farther research. In addition to the earlier agreement with Einstein Field Equations in curved spacetime, by imposing condition (\ref{normCond}) on normalized stress-energy tensor (\ref{CH2}) (hereinafter referred to as Alena Tensor) in flat Minkowski spacetime, one obtains consistent results, developing the knowledge of the physical system with electromagnetic field. Gravitational, electromagnetic and sum of other forces acting in the system may be expressed as shown in (\ref{forceDenNew}) where gravitational four-force density is dependent on the pressure $p$ in the system and equal to $f^{\mu}_{gr} = \varrho \left ( \mds{\, \ln{(p)}} U^{\mu} - c^2 \partial^{\mu} \ln{(p)} \right )$.\\ 
~\\
The conclusions from the article can be divided according to their areas of application, as conclusions for QED, conclusions for QM and conclusions regarding the combination of QFT with GR.

\subsection{Conclusions for QED}

Condition (\ref{normCond}) leads to electromagnetic four-potential for which some gauge may be expressed as $\mathbb{A}^{\mu} = - \frac{\Lambda_{\rho}}{p} \frac{\varrho_o}{\rho_o} U^{\mu}$. It simplifies Alena Tensor (\ref{CH2}) to familiar form $T^{\alpha \beta} = \frac{1}{\mu_o}\mathbb{F}^{\alpha \gamma} \partial^{\beta} \mathbb{A}_{\gamma} -  \Lambda_{\rho} \eta^{\alpha \beta}$ and both Lagrangianand and Hamiltonian density for the system with electromagnetic field appear to be related to invariant of the electromagnetic field tensor $\mathcal{L} = \mathcal{H} = \Lambda_{\rho} = \frac{1}{4\mu_o} \mathbb{F}^{\alpha \beta} \mathbb{F}_{\alpha \beta} = \frac{1}{2 \mu_o}\mathbb{F}^{0 \gamma} \partial^{0} \mathbb{A}_{\gamma}$.\\
~\\
The above would also simplify the Lagrangian density used in QED. Assuming that there is only electromagnetic field in the system and substituting $\Lambda_{\rho}$ for the current Lagrangian density used in QED, one should obtain equations that describe the entire system with electromagnetic field. Interestingly, such equations would also take into account the behavior of the system related to gravity, because according to the model presented here, gravity naturally arises in the system as a consequence of the existence of other fields (more precisely - existence of the energy momentum tensors associated with these fields) and the resulting Lagrangian density takes this into account.\\
~\\
Perhaps, this could explain why it is so difficult to find quantum gravity as a separate interaction within QFT, and, which is also possible, could also explain QED's extremely accurate predictions assuming that it actually describes the entire system with an electromagnetic field.

\subsection{Conclusions for QM}

As was shown in the article, $H^{\beta} \equiv - \frac{1}{c} \int T^{0 \beta} \, d^3 x$ acts as canonical four-momentum for the point-like particle and the vanishing four-divergence of $H^{\beta}$ turns out to be the consequence of Poynting theorem.\\
~\\
Obtained canonical four-momentum $H^{\mu}$ should satisfy the Klein-Gordon equation (\ref{Q10}) and it is equal to  
\begin{equation} \label{Q2}
H^{\mu} = P^{\mu} + V^{\mu} = - \frac{\gamma L}{c^2}U^{\mu} + \mathbb{S}^{\mu}
\end{equation}
where $P^{\mu}$ is four-momentum, L is Lagrangian for point-like particle, $\mathbb{S}^{\mu}$ due to its properties, seems to be some description of the spin, and where $V^{\mu}$ describes the transport of energy due to the field. It may be calculated as
\begin{equation} \label{QV1}
V^{\mu} = q \mathbb{A}^{\mu} + \frac{\varrho c^2 \gamma^2}{p} P^{\beta} + \frac{\varrho c^2}{p} \mathbb{S}^{\mu} + Y^{\mu} 
\end{equation}
where $\mathbb{A}^{\mu}$ is electromagnetic four-potential and where $Y^{\mu}$ is the volume integral of the Poyinting four-vector. \\
~\\
It seems that in such approach it would be possible to isolate gravity as a separate interaction, although this would probably require further research on the influence of individual components on the behavior of the particle. It is also not obvious how to deal with the interpretation of time in first quantization, however a clue may be to rely on possibility of using Geroch's splitting \cite{GEROCH} providing (3+1) decomposition.

\subsection{Conclusions regarding the combination of QFT with GR}

It should be noted that the presented solution applies to a system with an electromagnetic field, but it allows for the introduction of additional fields while maintaining the properties of the considered Alena Tensor. Therefore, it seems a natural direction for further research to verify how the system with additional fields will behave and what fields are necessary to obtain the known configuration of elementary particles and interactions.\\
~\\
For example, remaining with the previous notation, one may describe the field in the system by some generalized field tensor $\mathbb{W}^{\alpha \beta \gamma}$ part of which is the electromagnetic field (e.g. electroweak). Such description provides more degrees of freedom compared to the simple example for electromagnetism from chapter (\ref{ShortSummary}), and allows to represent the Alena Tensor in flat spacetime as follows
\begin{equation} \label{GEN1}
T^{\alpha \beta} =  \varrho \, U^{\alpha} U^{\beta} - \left ( \frac{c^2 \varrho}{\Lambda_{\rho}} +  1  \right ) \left ( \Lambda_{\rho} \, \eta^{\alpha \beta} - \mathbb{W}^{\alpha \delta \gamma} \, \mathbb{W}^{ \beta}_{ \; \delta  \gamma} \right )  
\end{equation}
where 
\begin{equation} \label{GEN2}
\Lambda_{\rho} \equiv \frac{1}{4} \mathbb{W}^{\alpha \beta \gamma} \mathbb{W}_{\alpha \beta \gamma}
\end{equation}
\begin{equation} \label{GEN3}
\xi \, h^{\alpha \beta} \equiv \frac{\mathbb{W}^{\alpha \delta \gamma} \, \mathbb{W}^{ \beta}_{ \; \delta  \gamma}}{\Lambda_{\rho}}
\end{equation}
\begin{equation} \label{GEN4}
\xi \equiv \frac{4}{\eta_{\alpha \beta} \, h^{\alpha \beta } }
\end{equation}
The Alena Tensor defined in this way retains most of properties described in the previous chapters, however, it now describes other four-force densities in the system, related to its vanishing four-divergence.\\
~\\
Further analysis of above using the variational method may lead to next discoveries regarding both the theoretical description of quantum fields and elementary particles associated with them, and the possibility of experimental verification of the obtained results.\\

\section{Statements} \label{4Chap}

Data sharing is not applicable to this article, as no datasets were generated or analyzed during the current study.\\
~\\
The author did not receive support from any organization for the submitted work.\\
~\\
The author has no relevant financial or non-financial interests to disclose.\\

\section*{References}

\bibliography{DevMethodAPS}

\end{document}